\documentclass[twocolumn,conference]{IEEEtran}
  \usepackage{subfigure}
  \usepackage{graphicx}
  \usepackage{makeidx}
  \usepackage{amsmath}
  \usepackage{amsfonts}
  \usepackage{bbding}
  \usepackage{amsfonts}
  \usepackage{amssymb}
  \usepackage{wrapfig}
  \usepackage{psfrag}
  \usepackage{epstopdf}
  \usepackage{cite}
  \usepackage[font=small,labelsep=period]{caption}
  \usepackage{bm}
  \usepackage{float}
  \usepackage{threeparttable}
  \usepackage{color}
  \usepackage{bbding}
  \usepackage{subfig}
  \usepackage{array,color}
  \usepackage{colortbl}
  \usepackage{amsfonts,amssymb}
  \usepackage{CJK}
  \usepackage{booktabs}
  \usepackage{multirow}
  \usepackage{url}
  \usepackage{stfloats}

  \ifCLASSINFOpdf
  \else
  \fi

\begin{document}

  \title{Blind Estimation Algorithms for I/Q Imbalance \\ in Direct Down-conversion Receivers}

  \author{
    \IEEEauthorblockN{Peiyang Song\IEEEauthorrefmark{1}\IEEEauthorrefmark{2}, Nan Zhang\IEEEauthorrefmark{1},  Hang Zhang\IEEEauthorrefmark{2} and Fengkui Gong\IEEEauthorrefmark{1}}

    \IEEEauthorblockA{\IEEEauthorrefmark{1}State Key Laboratory of ISN, Xidian University, Xi'an 710071, China}
     \IEEEauthorblockA{\IEEEauthorrefmark{2}Science and Technology on Communication Networks Laboratory, Shijiazhuang, China}
     \IEEEauthorblockA{Email: {pysong@stu.xidian.edu.cn, nzhang@xidian.edu.cn, 80090385@qq.com, fkgong@xidian.edu.cn}}
   }


  \maketitle

  \begin{abstract}
  As known, receivers with in-phase and quadrature-phase (I/Q) down conversion, especially direct-conversion architectures, always suffer from I/Q imbalance. I/Q imbalance is caused by amplitude and phase mismatch between I/Q paths. The performance degradation resulting from I/Q imbalance can not be mitigated with simply   higher signal to noise ratio (SNR). Thus, I/Q imbalance compensation in digital domain is critical. There are two main contributions in this paper. Firstly, we proposed a blind estimation algorithm for I/Q imbalance parameters based on joint first and second order statistics (FSS) which has a lower complexity than conventional Gaussian maximum likelihood estimation (GMLE). This can be used for further precessing such as equalization in presence of receiver IQ imbalance. In addition, we find out the reason of error floor in conventional I/Q imbalance compensation method based on conjugate signal model (CSM). The proposed joint first order statistics and conjugate signal model (FSCSM) compensation algorithm can reach the ideal bit error rate (BER) performance.
  \end{abstract}

  \begin{IEEEkeywords}
 I/Q imbalance, blind estimation algorithm, first and second order statistics, conjugate signal model
  \end{IEEEkeywords}

  \section{Introduction}

  \IEEEPARstart{W}{ith} increasingly demanding requirements for low-cost and low-power wireless receivers, I/Q imbalance problem has attracted more attention in both industrial and academic communities. I/Q imbalance is caused by amplitude and phase mismatch between in-phase and quadrature-phase branches in analog front-end, which is unavoidable in practical implementation. It always leads to serious performance degradation in receivers, especially those with low-cost radio frequency front-end (e.g. direct conversion architecture). This distortion, unfortunately, can not be mitigated by simply improving SNR. Hence, dealing with I/Q imbalance in digital domain becomes critical, especially for communications with large modulation order.

  Most literatures so far focus on data-aided estimation algorithms, such as least mean squares (LMS) \cite{lms_method}, decision-directed (DD) \cite{dd_method}, the expectation maximization (EM) \cite{em_based}, maximum likelihood (ML) \cite{ml_method}, minimum mean squared error (MMSE) \cite{mmse_method}, etc. More recently, \cite{iterative} describes an iterative I/Q compensation algorithm using both the training symbols and data symbols. Estimation by known symbols usually achieves a rapid convergence. However, the pilot symbols can also result in low spectral efficiency. This problem has become particularly severe as the spectrum resources are increasingly valuable.

  A few effective blind algorithms can also be found for I/Q imbalance estimation. \cite{advanced_method}, for the first time, introduces the blind source separation (BSS) into I/Q imbalance estimation. The advanced BSS technology, like joint approximative diagonalization of eigenmatrix (JADE) algorithm \cite{lianheduijiaohua}, is an efficient approach to correct I/Q imbalance, whereas it always suffers from significantly higher complexity and severe performance degradation when interfered by frequency offset. \cite{blind_signal} treats the I/Q imbalance problem as a conjugate signal model (CSM), where the observed signal is a linear combination of the desired signal and its complex conjugate. However, the algorithm always suffers from error floor which is especially severe in high order modulation. In addition, \cite{mmse_wl} and \cite{mmse_tsp} derive estimated values of amplitude and phase mismatch by approximating quadrature amplitude modulation (QAM) symbols as two-dimensional Gaussian variables. The estimated parameters are used for equalization in presence of receiver I/Q imbalance. Especially, \cite{jointblind} Addresses the compensation of transmitter I/Q imbalances and carrier frequency offset (CFO) for uplink single-carrier interleaved frequency-division multiple-access (SC-IFDMA) systems, which is of greater interest for the 5G networks.


  Throughout the letter, we define the notation as follows. We use bold-face upper case letters like $\bm  X$ to denote matrices, bold-face lower case letters like $\bm  x$ to denote column vectors, and light-face italic letters like $x$ to denote scalers. $x_i$ is the $i$th element of vector $\bm x$. $\bm x_i$ is the $i$th column vector of matrix $\bm X$. The complex conjugate of a complex $x$ is represented as $x^*$. $\bm I$ is the identity matrix. $\Re\{\cdot\}$ and $\Im\{\cdot\}$ denote the real and imaginary part of complex numbers, respectively. $\mathbb E[\cdot]$ represents the expectation. $[\cdot]^T$ and $[\cdot]^H$ denote the transpose and conjugate transpose operations, respectively. $\hat x$ (or $\bm {\hat X}$) is the estimated value of $x$ (or $\bm X$).

  \section{I/Q Imbalance System Model}
   Mathematical model of I/Q imbalance has been established and widely used \cite{advanced_method, blind_signal,jointblind,dd_method,em_based,high_order,iterative,lianheduijiaohua,lms_method,ml_method,mmse_method,mmse_tsp,mmse_wl,neural_networks}. In this section, we consider the transmission of baseband signals over a flat-frequency noisy channel and I/Q imbalance in direct conversion receivers. Bandpass signal is given by
  \begin{equation}  \label{rf_signal}
  x_{RF}(t)=I(t)\mbox{cos}(2\pi f_ct)-Q(t)\mbox{sin}(2\pi f_ct),
  \end{equation}
  where $f_c$ is the carrier frequency, $I(t)$ is the in-phase component of baseband signal and $Q(t)$ is the quadrature-phase one. The received signal can be written as
  \begin{equation}
  \begin{split}
  r(t)&=x_{RF}(t)+\omega_{RF}(t)  \\
  &=r_c(t)\mbox{cos}(2\pi f_c t)-r_s(t)\mbox{sin}(2\pi f_c t) ,
  \end{split}
  \end{equation}
  where $\omega_{RF}(t)$ is real-valued additive white Gaussian noise (AWGN). $r_c(t)$ and $r_s(t)$ are equivalent baseband signals which contain transmitted signals and the noise.
  In direct conversion architecture, down conversion is realized by multiplying received signal $r_{RF}(t)$ by local carrier and then passing the result through a low-pass filter. As described before, the in-phase local oscillator (LO) signal $z_{LO,c}(t)$ and the quadrature-phase one $z_{LO,s}(t)$ always exhibit both amplitude and phase mismatch, i.e.
  \begin{subequations} \label{lo_signal}
  \begin{align}
      z_{LO,c}(t)&=2(1+\alpha)\mbox{cos}(2\pi f_c t+\theta),  \\
      z_{LO,s}(t)&=-2(1-\alpha)\mbox{sin}(2\pi f_c t-\theta),
  \end{align}
\end{subequations}
  where $\alpha$ and $\theta$ are amplitude and phase imbalance parameters, respectively. It is worth noting that amplitude mismatches in different branches do not need to be exactly equal. The symbol $\alpha$ is used to make the formula more symmetrical. Thus, the final output baseband signal influenced by I/Q imbalance in receiver can be written as
  \begin{subequations}
  \begin{equation} \label{real_iq_model1}
  \begin{split}
  y_c(t)&=LPF\{z_{LO,c}(t)r(t)\}  \\
  &=\underbrace {(1+\alpha)\mbox{cos}(\theta)r_c(t)}_{signal}+\underbrace {(1+\alpha)\mbox{sin}(\theta)r_s(t)}_{interference} ,
  \end{split}
  \end{equation}
  \begin{equation} \label{real_iq_model2}
  \begin{split}
  y_s(t)&=LPF\{z_{LO,s}(t)r(t)\}  \\
  &=\underbrace {(1-\alpha)\mbox{sin}(\theta)r_c(t)}_{interference} +\underbrace {(1-\alpha)\mbox{cos}(\theta)r_s(t)}_{signal} ,
  \end{split}
  \end{equation}
\end{subequations}
  where $LPF\{\cdot\}$ accounts for low-pass filter. $y_c(t)$ and $y_s(t)$ are baseband signals of I/Q paths. As can be seen from (\ref{real_iq_model1}) and (\ref{real_iq_model2}), when ideal LO is taken into consideration (i.e. $\alpha=0$ and $\theta=0$), receivers can successfully recover the original baseband signals as $y_c(t)=r_c(t)$ and $y_s(t)=r_s(t)$. But in practice, when amplitude and phase imbalance occur, each of these two branches will be interfered by the other. (\ref{real_iq_model1}) and (\ref{real_iq_model2}) can be expressed by $\bm {y(t)}=[y_c(t) \quad y_s(t)]^T$ as
  \begin{equation} \label{matrix_iq_model}
  \bm {y(t)}
   =\underbrace{\left[ \begin{matrix}
   (1+\alpha)\mbox{cos}(\theta) & (1+\alpha)\mbox{sin}(\theta)  \\
   (1-\alpha)\mbox{sin}(\theta) & (1-\alpha)\mbox{cos}(\theta)
   \end{matrix} \right]}_{\bm \Gamma}
   \left[ \begin{matrix}
   r_c(t) \\
   r_s(t)
  \end{matrix} \right].
  \end{equation}

\begin{figure}[h]
    \centering
    \includegraphics[width=180pt]{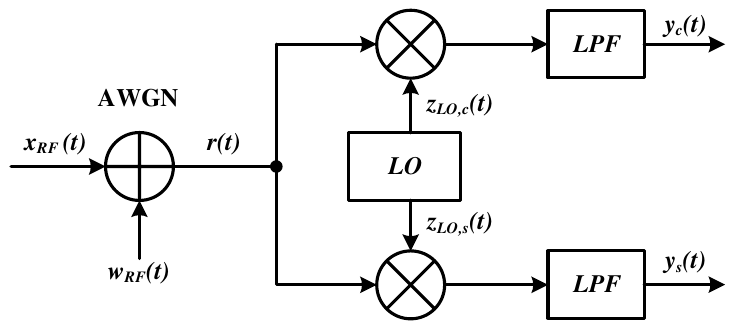}
    \caption{I/Q imbalance system model with real signals}
    \label{first_block}
  \end{figure}

  Fig. \ref{first_block} is the block diagram of this system model.

  The model can be derived by the complex envelope \cite{wireless_comm} of real signal as well. Received signal $r(t)$ can be given as
  \begin{equation}
  \begin{split}
  r(t)&=\Re\left\{ \widetilde r(t) e^{j2\pi f_c t}\right\}  \\
  &=\frac{1}{2} \left(\widetilde r(t)e^{j2\pi f_c t}+\widetilde r^*(t)e^{-j2\pi f_c t}\right),
  \end{split}
  \end{equation}
  where $\widetilde r(t)=r_c(t) + j r_s(t)$ is the complex envelope of received signals $r(t)$. Here, local carrier generated by LO can be written as
  \begin{equation}
  \widetilde z_{LO}(t)=2\left(K_1e^{-j2\pi f_c t}+K_2e^{j2\pi f_c t}\right),
  \end{equation}
  where
  \begin{subequations}
  \begin{align}
  K_1&=[(1-\alpha)e^{j\theta} +(1+\alpha)e^{-j\theta}] /2 ,  \\
  K_2&=[(1+\alpha)e^{j\theta} -(1-\alpha)e^{-j\theta}] /2.
  \end{align}
\end{subequations}

\begin{figure}[h]
   \centering
   \includegraphics[width=200pt]{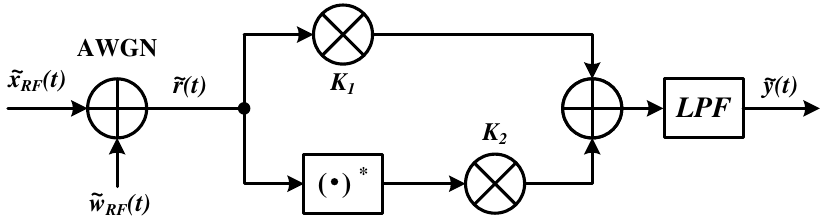}
   \caption{I/Q imbalance system model with equivalent complex envelope, where $(\cdot)^*$ denotes conjugate value.}
   \label{second_block}
 \end{figure}

  Equivalent complex signal after down conversion can be given as
  \begin{equation} \label{conjugate_iq_model}
  \widetilde y(t)=LPF\{r(t) \widetilde z_{LO}(t)\}=\underbrace {K_1 \widetilde r(t)}_{signal} + \underbrace {K_2 \widetilde r^*(t)}_{interference} ,
  \end{equation}
  which is shown in Fig. \ref{second_block}.

  As can be seen from (\ref{conjugate_iq_model}), when ideal LO is implemented, $K_1=1$ and $K_2=0$. Baseband signal can be accurately recovered without interference. If $\alpha$ and $\theta$ can not be ignored, the entire complex baseband signal will be interfered by the conjugate value of itself.

  Obviously, (\ref{conjugate_iq_model}) is equivalent to (\ref{real_iq_model1}) and (\ref{real_iq_model2}), i.e.
  \begin{equation}
  y_c(t)=\Re \{{\widetilde y(t)}\} \quad \mbox{and} \quad y_s(t)=\Im \{{\widetilde y(t)\}}.
  \end{equation}

  Then, we consider the QAM modulation and assume the in-phase and quadrature-phase components of the transmitted signal as
  \begin{subequations}
  \begin{align}
  I(t)=\sum\nolimits_{\ell} a_{\ell} \psi(t-{\ell}T) ,  \\
  Q(t)=\sum\nolimits_{\ell} b_{\ell} \psi(t-{\ell}T) ,
  \end{align}
\end{subequations}
  where, $a_\ell$ and $b_\ell$ belong to the QAM alphabet. $\psi(t)$ is a square-root raised-cosine pulse, and $T$ is the symbol period. The discrete-time in-phase and quadrature-phase components after down-conversion and matched filtering can be written as
  \begin{subequations}
  \begin{equation} \label{basic_model1}
  \begin{split}
  & y_{c,k} = y_c(kT)  \\
  &=(1+\alpha)[(a_k+n_{c,k})\mbox{cos}(\theta)+(b_k+n_{s,k})\mbox{sin}(\theta)] , \\
  \end{split}
  \end{equation}
  \begin{equation}  \label{baisc_model2}
  \begin{split}
  & y_{s,k}=y_s(kT)  \\
  &=(1-\alpha)[(a_k+n_{c,k})\mbox{sin}(\theta)+(b_k+n_{s,k})\mbox{cos}(\theta)] ,
  \end{split}
  \end{equation}
\end{subequations}
  where $n_{c,k}$ and $n_{s,k}$, which uncorrelate with each other, are the samples of zero-mean real-valued Gaussian random variables with variance $\sigma_n^2/2$.

    \begin{figure*}[b]
      \hrulefill
    \begin{equation} \label{con_zero}
      \tag{14}
      \bm C_\Omega= \mathbb E\left[\bm \Omega \bm \Omega^T\right]
      =\left[ \begin{matrix}
      \mbox{cos}^2(\theta) \mathbb E[r_{c,k}^2]+\mbox{sin}^2(\theta)  \mathbb E[r_{s,k}^2]  &  \mbox{sin}(\theta) \mbox{cos}(\theta) ( \mathbb E[r_{c,k}^2]+ \mathbb E[r_{s,k}^2])  \\
      \mbox{sin}(\theta) \mbox{cos}(\theta)( \mathbb E[r_{c,k}^2]+ \mathbb E[r_{s,k}^2])  &  \mbox{sin}^2(\theta) \mathbb E[r_{c,k}^2]+\mbox{cos}^2 (\theta) \mathbb E[r_{s,k}^2]
      \end{matrix} \right]
      \end{equation}
  \end{figure*}

  \section{Proposed joint first and second order statistics (FSS) algorithm}
  Firstly, $\alpha$ and $\theta$ are set as the trial values of I/Q imbalance parameters. \cite{mmse_wl} proposed an estimation algorithm for $\alpha$ and $\theta$ by Gaussian maximum likelihood. The estimated parameters are used for equalization in presence of I/Q imbalance.
  Here, we propose a simpler algorithm for estimation of $\alpha$ and $\theta$. It is easy to break (\ref{basic_model1}) and (\ref{baisc_model2}) into
  \begin{equation} \label{hahahaha}
  \bm Y=
  \left[\begin{matrix}
    \bm y_c \\ \bm y_s
  \end{matrix}\right]
  =
  \underbrace{
  \left[ \begin{matrix}
  1+\alpha & 0 \\ 0 & 1-\alpha
  \end{matrix} \right]
  }_{\bm \Psi}
  \underbrace{
  \left[ \begin{matrix}
  \mbox{cos}(\theta) & \mbox{sin}(\theta) \\ \mbox{sin}(\theta) & \mbox{cos}(\theta)
  \end{matrix} \right]
  \left[ \begin{matrix}
  \bm r_c \\ \bm r_s
  \end{matrix} \right]
  }_{\bm \Omega} ,
  \end{equation}
  where $r_{c,k}=a_k+n_{c,k}$ and $r_{s,k}=b_k+n_{s,k}$ are samples of $r_c(t)$ and $r_s(t)$ which contain both QAM symbols and noise. (\ref{hahahaha}) can be written as $\bm Y=\bm \Gamma \bm R$, where $\bm R=[\bm r_c, \bm r_s]^T$. When $\alpha$ and $\theta$ have been estimated, original signal can be recovered by $\bm R=\bm \Gamma^{-1} \bm Y$.

  We use $\bm C_R$, $\bm C_\Omega$ and $\bm C_Y$ represent covariance matrix of $\bm R$, $\bm \Omega$ and $\bm Y$ respectively. With $\mathbb E[\bm \Omega_k]=[0, 0]^T$, $\bm C_\Omega$ can be given as (\ref{con_zero}) (see the bottom of this page).

  Then, we assume that the received signals before down conversion are circular symmetry, i.e. $ \mathbb E[(r_{c,k}+jr_{s,k})^2] = 0$ , which can be further written as $\mathbb E[r_{c,k}^2]=\mathbb E[r_{s,k}^2]$ and $\mathbb E[r_{c,k}r_{s,k}]=0$.

  We set $P_r=2\mathbb E[r_{c,k}^2]=2\mathbb E[r_{s,k}]^2$ as the average power of received symbols. $\bm C_R$ and $\bm C_\Omega$ can be written as
  \setcounter {equation} {14}
  \begin{subequations}
  \begin{align} \label{c_omega}
  \bm C_R &= \frac{1}{2}\mathcal P_r \bm {\bm I} ,  \\
  \bm C_\Omega &= \frac{1}{2}\mathcal P_r \left[ \begin{matrix}
  1 & \mbox{sin}(2\theta)  \\
  \mbox{sin}(2\theta) & 1
  \end{matrix} \right] ,
  \end{align}
\end{subequations}
  which reveals that phase mismatch $\theta$ will not change the average power (i.e. the variance) of received symbols of each branch. Similar to $\bm C_\Omega$, $\bm C_{\bm Y}$ can be written as
  \begin{equation} \label{c_y}
  \begin{split}
  \bm C_{\bm Y} &= \mathbb E[\bm \Psi \bm \Omega \bm \Omega ^T \bm \Psi ^T] = \mathbb E[\bm \Psi \bm C_\Omega \bm \Psi^T]  \\
  &=\frac{1}{2}\mathcal P_r \left[ \begin{matrix}
  (1+\alpha)^2  &  (1-\alpha^2)\mbox{sin}(2\theta) \\
  (1-\alpha^2)\mbox{sin}(2\theta)  & (1-\alpha)^2
  \end{matrix} \right] .
  \end{split}
  \end{equation}

  To avoid power measurement, we give a calculation of $P_r$ by baseband signal $\bm y_s$ and $\bm y_c$. Resulting from the \emph{i.i.d}. property of $a_k$ and $b_k$ (also $n_{c,k}$ and $n_{s,k}$), it is obvious that
  \begin{equation}
    \begin{split}
    \eta &= \mathbb E[|(a_k+n_{c,k})\mbox{cos}(\theta)+(b_k+n_{s,k})\mbox{sin}(\theta)|]  \\
    &= \mathbb E[|(a_k+n_{c,k})\mbox{sin}(\theta)+(b_k+n_{s,k})\mbox{cos}(\theta)|] .
    \end{split}
  \end{equation}

 Thus, relationship between $\alpha$ and received symbols can be derived by (12a) and (12b) as
 \begin{equation} \label{alpha_tmp}
  \mathbb E[|y_{c,k}|]=(1+\alpha)\eta \quad \mbox{and} \quad \mathbb E[|y_{s,k}|]=(1-\alpha)\eta  .
 \end{equation}

 Then, we replace the statistical average by actual received symbols. And (\ref{alpha_tmp}) can be written as
 \begin{equation}
  \frac{1+\alpha}{1-\alpha}=\frac{\sum_{k=1}^N |y_{c,k}|}{\sum_{k=1}^N|y_{s,k}|} ,
 \end{equation}
  where $\alpha$ can be easily calculated by received symbols. Also, we give the expression of $\theta$ from (\ref{c_y}) as
  \begin{equation} \label{theta_tmp}
  \frac{1}{2}(1-\alpha^2)\mathcal P_r \mbox{sin}(2\theta)=\mathbb E[y_{c,k}y_{s,k}]=\mathbb E[y_{s,k}y_{c,k}] .
  \end{equation}

  Then, we replace the statistical average by actual received symbols. And (\ref{theta_tmp}) can be written as
  \begin{equation}
    \frac{1}{2}(1-\alpha^2)\mathcal P_r \mbox{sin}(2\theta)=\frac{1}{N}\sum\nolimits_{k=1}^N (y_{c,k}y_{s,k}) ,
  \end{equation}
  where $\mathcal P_r$ can be obtained from (\ref{c_y}) by received symbols and calculated $\alpha$ as
  \begin{equation}
    \mathcal P_r = \frac{1}{N} \left[\frac{\sum_{k=1}^N y_{c,k}^2}{(1+\alpha)^2} + \frac{\sum_{k=1}^N y_{s,k}^2}{(1-\alpha)^2}\right] .
  \end{equation}

  At last, $\alpha$ and $\theta$ can be written as
  \begin{subequations}
  \begin{align} \label{first_result}
    \alpha&=\frac{\sum_{k=1}^N |y_{c,k}|-\sum_{k=1}^N |y_{s,k}|}{\sum_{k=1}^N |y_{c,k}|+\sum_{k=1}^N |y_{s,k}|} ,  \\
    \theta=\frac{1}{2} &\mbox{arcsin}\left[
      \frac{2(1-\alpha^2)\rho_{cs}}
      {(1-\alpha)^2\rho_c + (1+\alpha)^2\rho_s}
    \right] ,
  \end{align}
\end{subequations}
where $\rho_{cs}=\sum\nolimits_{k=1}^N (y_{c,k}y_{s,k})$, $\rho_c=\sum\nolimits_{k=1}^N y_{c,k}^2$ and $\rho_s=\sum\nolimits_{k=1}^N y_{s,k}^2$.

As known, frequency offset does not change the circular symmetry property of received signal. Hence, the proposed algorithm can achieve a strong robustness to frequency offset.


\section{Proposed joint first order statistics and conjugate signal model (FSCSM) algorithm}
Firstly, we give a brief introduction of conventional CSM algorithm. The blind I/Q imbalance estimation is addressed in \cite{blind_signal} with conjugate signal model, i.e.
\begin{equation}
\widetilde {\bm Y} =
\left[ \begin{matrix}
\widetilde {\bm y} \\ \widetilde {\bm y}^*
\end{matrix} \right]
=\left[ \begin{matrix}
K_1 & K_2  \\ K_2^* & K_1^*
\end{matrix} \right]
\left[\begin{matrix}
\widetilde {\bm r} \\ \widetilde {\bm r}^*
\end{matrix}\right]
= \bm K \widetilde {\bm R} .
\end{equation}

Assuming that the target signal $\widetilde {\bm r}$ is circular or proper, the target here is to find a matrix $\bm W$ to whiten or decorrelate the components of $\widetilde {\bm r}$ as
\begin{equation}
\widetilde {\bm z} = \bm W \widetilde {\bm Y} = \bm W \bm K \widetilde {\bm R} = \bm T \widetilde {\bm R} ,
\end{equation}
where $\bm T$ is the equivalent system matrix. When perfect estimation is taken into consideration, $\bm T\approx \bm I$. $\bm W$ can be calculated as
\begin{equation} \label{secondary_batch}
\bm W=\bm U \bm \Lambda ^{-1/2} \bm U^H ,
\end{equation}
where $\bm U$ and $\bm \Lambda$ are calculated by eigenvalue decomposition of $\bm C_{\widetilde {\bm Y}}$, i.e.
\begin{equation}
\bm C_{\widetilde {\bm Y}} = \mathbb E (\widetilde {\bm Y} \widetilde {\bm Y}^H) = \bm U \bm \Lambda \bm U^H .
\end{equation}

Compared to earlier BSS technology, CSM algorithm has a much lower complexity. However, its performance seems not to be reliable enough. In practical, we find that CSM algorithm always suffers from a certain error floor, which is especially severe in high order modulations. A large number of simulations revealed that the error floor usually mitigates as the amplitude mismatch $\alpha$ reduces. When $\alpha$ is not taken into consideration, BER performance of CSM algorithm can be improved to ideal bound.

(\ref{hahahaha}) has proved that the influence of $\alpha$ and $\theta$ on impaired signals is independent. Thus, one possible solution to eliminate the error floor is compensating amplitude mismatch in advance. Our proposed FSCSM algorithm add a pre-processing module before conventional CSM method. The pre-processing can be given as

\begin{equation} \label{sosa_csm}
  \begin{split}
\bm Y'&=\left[\begin{matrix} \bm y'_c \\ \bm y'_s \end{matrix}\right]=
    \hat {\bm \Psi}^{-1} \bm Y =\hat {\bm \Psi}^{-1} \bm \Psi \bm \Omega
    = \left[ \begin{matrix}
      \frac{1}{1+\hat \alpha} & 0 \\ 0 & \frac{1}{1-\hat \alpha}
      \end{matrix} \right]
      \left[ \begin{matrix}
      \bm y_{c}  \\ \bm y_{s}
      \end{matrix} \right]  \\
    &= \underbrace{
      \left[ \begin{matrix}
      \frac{1+\alpha}{1+\hat \alpha} & 0 \\ 0 & \frac{1-\alpha}{1-\hat\alpha}
      \end{matrix} \right]
      }_{\bm \Psi'}
      \underbrace{
      \left[ \begin{matrix}
      \mbox{cos}(\theta) & \mbox{sin}(\theta) \\ \mbox{sin}(\theta) & \mbox{cos}(\theta)
      \end{matrix} \right]
      \left[ \begin{matrix}
      \bm r_c \\ \bm r_s
      \end{matrix} \right]
      }_{\bm \Omega},
    \end{split}
\end{equation}
where $\hat \alpha$ is the estimated value of amplitude mismatch calculated by (23a). $\bm \Psi'=\hat {\bm \Psi}^{-1} \bm \Psi$ is the remaining amplitude mismatch matrix. When accurate estimation of $\alpha$ is performed, $\bm \Psi'\approx\bm I$ and $\bm Y'\approx\bm\Omega$. Then, the equivalent complex baseband signal for CSM is $\bm Y '=[\bm {\widetilde y'},\bm {\widetilde y'}^*]^T$, where $\bm {\widetilde y'}=\bm y'_c+j\bm y'_s$.

\begin{figure}[h]
  \centering
\includegraphics[width=230pt]{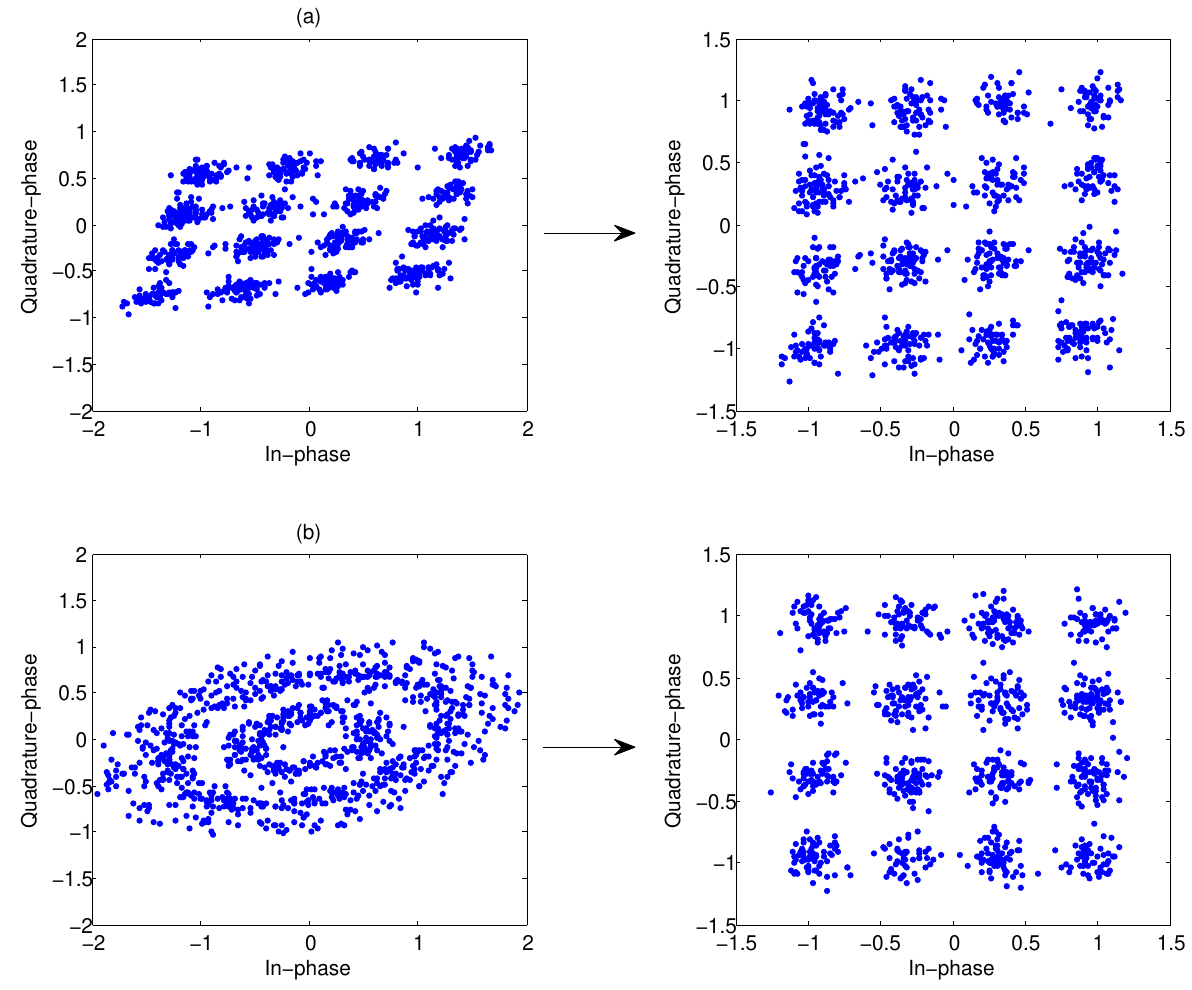}
\caption{Constellation comparison pre- and post- FSCSM algorithm. (a) signal is impaired by I/Q imbalance. (b) signal is impaired by both I/Q imbalance and frequency offset.}
\label{xingzuotu}
\end{figure}

The effect of proposed FSCSM algorithm can be shown in Fig. \ref{xingzuotu}. Signals impaired by I/Q imbalance can be perfectly compensated. Frequency offset will be completely reserved and wait for further processing. Here we consume that the frequency offset is known and has been fully correct after our FSCSM module. Further performance analysis will be given in Section \ref{numerical_results}.

  \section{Numerical Results} \label{numerical_results}
\begin{figure}[h]
      \centering
    \includegraphics[width=240pt]{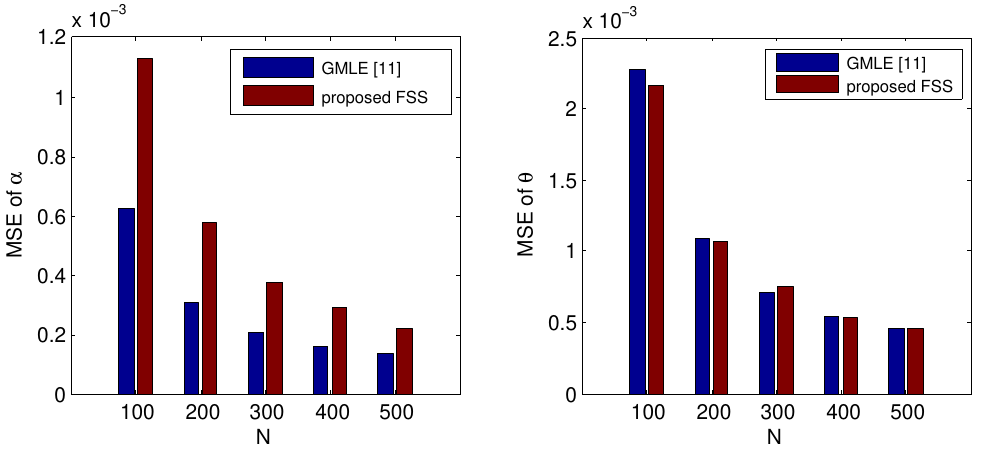}
    \caption{Mean square error (MSE) of estimated $\alpha$ and $\theta$ achieved by GMLE and proposed FSS}
    \label{tiaoxingtu}
    \end{figure}

  In this section, we assess the performance and robustness of our proposed algorithms. Table \ref{weiyibiaoge} shows the operations required for GMLE and proposed FSS. As can be seen, compared to GMLE, our proposed FSS estimation is free of complicated square root operations. Also, less additions and multiplications are required. Additional bit shift and absolute value operations introduced by proposed FSS are relatively easy for implementation. Fig. \ref{tiaoxingtu} reports the mean square error of estimated $\alpha$ and $\theta$ achieved by GMLE and FSS versus the number of received 16-QAM symbols $N$, for SNR=18dB. Estimation results of both algorithms are quite close which means the accuracy degradation of proposed FSS caused by simplification is slight enough.

  \begin{table*}[!ht]
    \renewcommand\arraystretch{1.5}
    \centering
    \captionsetup{font={small},format=plain,
    singlelinecheck=off, labelsep=newline,justification=centering}
    \caption{\textsc{Operation number required for GMLE and proposed FSS algorithms}}
    \label{weiyibiaoge}
    \begin{tabular}{cccccccc}
    \toprule
    Algorithm & Square Root & Multiplication & Addition & Absolute Value of Real Number & Bit Shift & Division & Arcsin \\
    \midrule
    GMLE \cite{mmse_tsp} & $3$ & $3N+2$ & $3N-1$ & $0$ & $0$ & $3$ & $1$ \\
    proposed FSS & $0$ & $6$ & $2N+4$ & $2N$ & $2$ & $3$ & $1$  \\
    \bottomrule
    \end{tabular}
    \end{table*}

\begin{figure}[h]
    \centering
  \includegraphics[width=240pt]{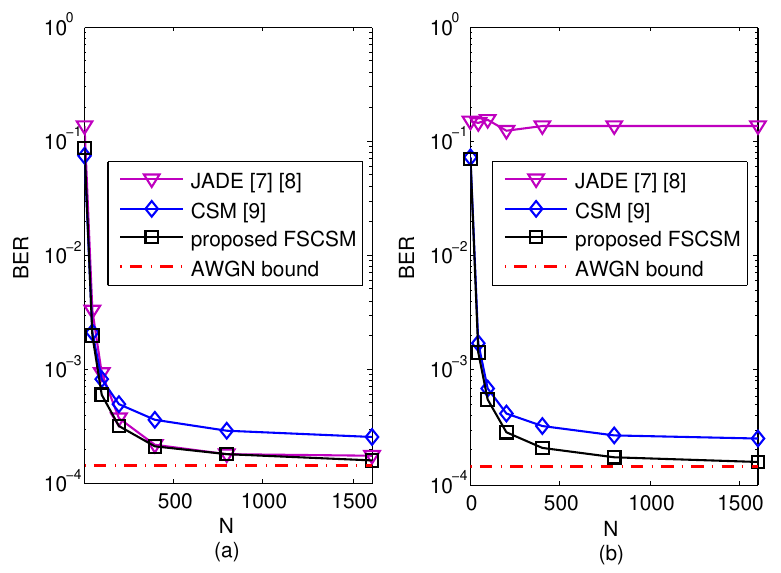}
  \caption{Convergence of different blind algorithms in 16-QAM, for SNR=18dB. (a) $\Delta f=0$; (b) $\Delta f=0.01$}
  \label{shoulianxing}
  \end{figure}

  Fig. \ref{shoulianxing} reports the convergence of different algorithms in 16-QAM modulation with SNR=18dB. I/Q imbalance is set as $\alpha=0.2$ and $\theta=10^\circ$. $\Delta f$ is the normalized CFO. To ignore the effect of frequency offset compensation algorithms on the BER performance, we assume that $\Delta f$ is known at receiver and is perfectly compensated after I/Q imbalance compensation.

  Simulation shows that our proposed  FSCSM algorithm performs a fast BER performance convergence. The BER can be eventually upgraded to AWGN bound (i.e. BER of ideal receiver without I/Q imbalance). Moreover, when frequency offset is taken into consideration, BER performance of both the proposed algorithms does not degrade.

\begin{figure}[h]
    \centering
  \includegraphics[width=210pt,height=170pt]{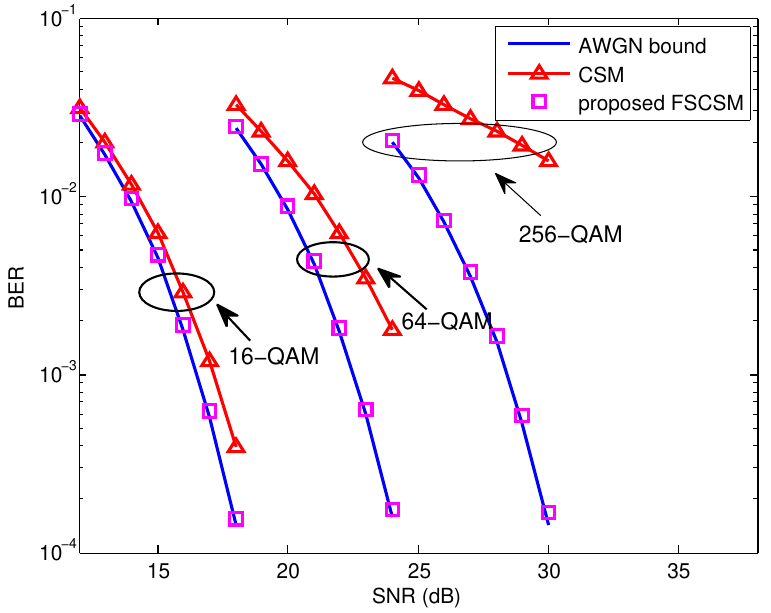}
  \caption{Performance of conventional CSM and proposed FSCSM algorithm in  different QAMs}
  \label{duibi}
  \end{figure}

  Fig. \ref{duibi} compares the performance of conventional CSM with proposed FSCSM blind algorithm in 16-QAM, 64-QAM and 256-QAM, respectively. Parameters are set as $\alpha=0.3$,  $\theta=10^\circ$ and $\Delta f=0$. And $N$ is large enough to guarantee that the BER performance has converged. As can be seen, for conventional CSM algorithm, the degradation of performance can not be mitigated by simply increase $N$. Our proposed FSCSM algorithm solves this problem without knowledge of any other additional information and improve the BER performance to AWGN bound.

  \section{Conclusion}
  This letter has addressed the I/Q imbalance problem in single-carrier direct conversion receivers. Two blind estimation and compensation algorithms have been put forward. The proposed FSS algorithm achieves less complexity than conventional GMLE estimation with slight enough performance degradation. In addition, the reason of error floor in conventional CSM algorithm is found out. Our proposed FSCSM algorithm can eliminate the error floor of CSM and achieve ideal BER performance. It is worth noting that the proposed two algorithms both perform a strong robustness to frequency offset, which makes it work well before frequency offset estimation algorithms. Furthermore, both the proposed algorithms are not sensitive to modulation order. Our results show that they also perform well with 4096-QAM and 256-APSK.

  \section*{Acknowledgement}

  This work is supported in part by joint fund of ministry of education of China (6141A02022338) and the opening project of science and technology on communication networks laboratory (KX162600027).

  \ifCLASSOPTIONcaptionsoff
    \newpage
  \fi

  \bibliographystyle{IEEETran}
  \bibliography{database.bib}

  \end{document}